\documentclass[10pt,prb,twocolumn,showpacs,superscriptaddress]{revtex4}

\usepackage{graphicx}

\newcommand{\be}{\begin{equation}}
\newcommand{\ee}{\end{equation}}
\newcommand{\bea}{\begin{eqnarray}}
\newcommand{\eea}{\end{eqnarray}}
\newcommand{\down}{\downarrow}
\newcommand{\up}{\uparrow}
\newcommand{\f}{\frac}

\newcommand{\Anu}{A^{\nu}_{q_y}}

\newcommand{\Bnu}{B^{\nu}_{q_x}}
\newcommand{\rhoonep}{\rho_{1 q_y}\left( q_x \right)}

\newcommand{\rhotwop}{\rho_{2 q_y}\left( q_x \right)}

\newcommand{\rhothrp}{\rho_{3 q_x}\left( q_y \right)}

\newcommand{\rhofoup}{\rho_{4 q_x}\left( q_y \right)}

\newcommand{\kper}{k^\perp}
\newcommand{\kpar}{k^{||}}
\newcommand{\cp}{(+)}
\newcommand{\cm}{(-)}
\newcommand{\dlU}{u}

\newcommand{\alpd}{\alpha^{\dagger q_y}_{q_x}}
\newcommand{\alp }{\alpha^{-q_y}_{-q_x}}
\newcommand{\betd}{\beta^{\dagger q_x}_{q_y}}
\newcommand{\bet }{\beta^{-q_x}_{-q_y}}
\newcommand{\ada}{a^{\dagger q_y}_{q_x}}
\newcommand{\alo}{a^{-q_y}_{-q_x}}
\newcommand{\bda}{b^{\dagger q_x}_{q_y}}
\newcommand{\blo}{b^{-q_x}_{-q_y}}

\newcommand{\VXY}{V_{q_x q_y}}
\newcommand{\VYX}{V_{q_y q_x}}

\begin{document}

\title{Adjacent face scattering of electrons on a square Fermi surface}

\author{Olav  F. Sylju{\aa}sen and A. Luther}
\affiliation{NORDITA, Blegdamsvej 17, DK-2100 Copenhagen {\O}, Denmark}
\email{sylju@nordita.dk}

\thanks{}

\date{\today}

\pacs{71.10 Fd}
\preprint{NORDITA-2004-93}

\begin{abstract}
Interacting electrons with a square Fermi surface is investigated from a bosonic
point of view taking into account electron scattering between all faces of the square.
Fermion operators are classified according to their dimensions and the stability of the boson fixed-point is investigated.  In particular we find, in contrast to previous studies, that
the square Fermi surface is unstable to doping in the case of no spin gap and microscopic Hubbard interactions.
\end{abstract}

\maketitle

\section{Introduction}
Much effort has been invested in solving the
two dimensional one-band Hubbard model which describes
repulsive electrons hopping on a square lattice.  The potential payoff
is a possible explanation of high-Tc superconductivity.
Yet no definite solution exists and it might be more valuable
to shift focus from the microscopic Hubbard model to more general
classes of effective models. Such a strategy is well formulated using
the renormalization group approach. The essence of this approach is that
low-energy properties of any microscopic Hamiltonian become identical, at low enough energies, to
the same properties of an effective Hamiltonian describing a fixed point
of the renormalization group flow. Instabilities and phase transitions can
be understood by the presence of relevant operators that cause flows away from this fixed point.

The best studied fixed point for 2D electrons is that of free electrons with a Fermi surface.
There are only a few instabilities associated with this fixed point\cite{Shankar}, among them the BCS superconducting instability. {\em Another} fixed point for 2D electrons can be arrived at by considering electrons on a square Fermi surface as a collection of electrons on coupled chains\cite{Luther}. 
This fixed-point Hamiltonian consists of the chain electron density operators as well
as forward-scattering terms and is purely bosonic.
In the context of 2D electrons this boson Hamiltonian has not been extensively studied 
except for the works \cite{Luther,Sudbo} where a simplifying assumption 
was employed: No interactions between density operators on adjacent faces of the square. 

In this article we extend the analysis of the boson Hamiltonian to also include adjacent face couplings in such a way that the most general interactions respecting the square symmetry
are taken into account. We diagonalize explicitly the boson Hamiltonian, and classify a family of fermion operators with respect to this fixed-point.  As will be shown, a particularly important consequence of adjacent face interactions is that  the stability of the square Fermi surface is reduced. 

It should be mentioned that the same boson Hamiltonian as considered here, with other values for the parameters, appears in the studies of the so called sliding Luttinger liquids\cite{Sliding}, both in the context of modeling striped phases of cuprates\cite{Kivelson} and the quantum Hall effect\cite{Fradkin}.

\section{The square Fermi surface mapping}
The construction used to map fermions on a square Fermi surface to chains was
explained in details in Ref.~\cite{Luther}. 
In the following paragraphs we will outline the essentials of this mapping.
The construction is similar in spirit to what is done in one dimension. There the real-space fermion operator is written as a sum of two terms, each
dominated by momentum components around one of the Fermi points. On the square Fermi surface, the fermion operator is decomposed into
a sum over the four faces of the square
\be
	\psi(x,y) = \sum_{p=1}^4 \psi_p(x,y)
\ee
where $\psi_p$ has momentum components coming from face $p$. This face field can be written in terms of momentum components, which have a natural decomposition into momentum components parallel and perpendicular
to the face normal. Identifying the momentum components having identical perpendicular momentum with
the momentum components of a one dimensional chain close to one of its Fermi points, one can
write the face field as a sum over real-space chains oriented along the face normal
\be \label{mapping}
    \psi_p(x,y) = \sqrt{\f{\pi}{k_f}} \sum_{l^\prime} g(l-l^\prime)
                  \psi_{l^\prime,p}(x)
\ee
where $k_f= \pi/(\sqrt{2}a)$, and
\be
   g(l) = \f{1}{2N} \sum_{n_y=-N/2}^{N/2} e^{i \pi n_y l/N} \stackrel{N\to \infty}{=} \f{\sin(\pi l/2)}{\pi l}
\ee
is the real space structure function coming from restricting the perpendicular momentum
coordinate which labels each chain, to lie within $[-k_f,k_f]$. We have rotated the coordinate system such that the integer label $l$ is related to the real space coordinate $y$ parallel to face 1, $\sqrt{2} y=a l$.
where $a$ is the lattice spacing.
Note that the square Fermi surface is used here as a bookkeeping tool only in order to keep
track of relevant states. Thus the real Fermi
surface of electrons does not necessarily need to be square.
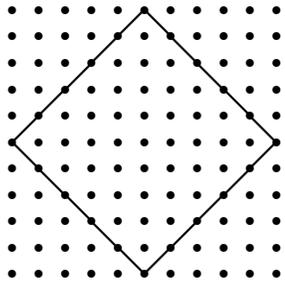
\begin{figure}
\begin{center}
\begin{picture}(100,100)
\multiput(0,0)(10,0){11}{\multiput(0,0)(0,10){11}{\circle*{3.0}}}
\thicklines
\put(50,0){\line(1,1){50}}
\put(100,50){\line(-1,1){50}}
\put(50,100){\line(-1,-1){50}}
\put(0,50){\line(1,-1){50}}
\end{picture}
\caption{The square Fermi surface \label{Square}}
\end{center}
\end{figure}

The Hamiltonian is obtained by first considering the kinetic energy term of free fermions with a
square Fermi surface. Linearizing the energy spectrum and using the mapping Eq.~(\ref{mapping}) one obtains
a bilinear term of fermion operators on in general different chains, but belonging to the same
face. Only those terms where the fermion operators belong to the same chain is retained in the fixed-point Hamiltonian,
thus the fixed-point Hamiltonian describes effectively a collection of 1D chains. This is bosonized
in the standard way giving a Hamiltonian quadratic in density operators for each chain. The effective
velocity $v_0$ associated is an average over the fermi velocity over the face.
The rest of the terms involving different chain operators are treated as perturbations to the boson fixed-point Hamiltonian.
There are also other terms corresponding to fermion-fermion interactions that can be written as a product of two chain density operators.
Using the mapping Eq.~(\ref{mapping}) for four-fermion interactions, pairwise chain contributions corresponding to pairs of density operators are also included in the fixed-point Hamiltonian, giving the fixed-point Hamiltonian properties of the interacting fermion system.

 The most general such Hamiltonian invariant under rotations quadratic in density operators is
\be
	H = \f{1}{2NL} \! \! \! \! \! \!  \sum_{
	\stackrel{p, p^\prime,\kpar_p,\kper_p}{\alpha=s,c}} \! \! \! \! \! \! \! \rho^\alpha_{p \kper_p}(\kpar_p)  A^\alpha_{pp^\prime}(\kper_p,\kper_{p^\prime}) \rho^\alpha_{p -\kper_p}(-\kpar_p) \label{genHam}
\ee 
where the labels $p$ and $p^\prime$ denotes the faces of the square and takes values from 1 to 4.
The momentum $\kpar_p$ is the momentum parallel to the face normal, that is along the chain and $\kper_p$ is the momentum along the Fermi surface, perpendicular to the chains.
We have divided the variables into charge and spin sectors: $\sqrt{2} \rho^c = \rho_\up+\rho_\down$
and $\sqrt{2} \rho^s = \rho_\up - \rho_\down$.
The matrix entries of $A$ are restricted by the rotational invariance. We have
$A_{ii}=A^+ + A^-$, $A_{ij}=A^+ - A^-$ for $i$ and $j$ labeling opposite phases, and
$A_{ij}=V$ for $i$ and $j$ labeling adjacent faces.

The contributions from the kinetic energy to these coupling constants are
\bea
	A^+_{\rm kin} & = & A^-_{\rm kin} = \f{\pi v_0}{2}   \nonumber \\
	V_{\rm kin} & = & 0 \nonumber
\eea
being the same in both the spin and the charge sector.
We will treat the Hamiltonian for generic couplings, but it is illustrative to
write down the explicit expressions in the case when the interaction term is the
Hubbard U term
\be
   H_U = \f{U}{2} \sum_{j} \left( n_{j \up} n_{j \down}+ n_{j \down} n_{j \up} \right).
\ee
Then the charge sector parameters take the form
\bea
	A^c_+ & = & \f{\pi v_0}{2} + \f{Ua}{2} f^2(\kper)   \nonumber \\
	A^c_- & = & \f{\pi v_0}{2}    \label{Hubbardparams} \\
	V^c   & = & \f{Ua}{2} f(\kpar) f(\kper)  \nonumber
\eea
where $f(k)$ is the Fourier transform of $g^2$ given by
\be
	f(k) = \sum_l g^2(l) e^{ikl} = \f{1}{2} \left( 1- \f{|k|}{\pi} \right).
\ee
The spin sector parameters are obtained by letting $U\to-U$.

\section{Diagonalization}
In order to diagonalize the Hamiltonian Eq.~(\ref{genHam}) we write it down explicitly.
It has the same form in the two ($\alpha=c,s$) sectors
\bea
  H_\nu & = & \f{1}{2NL} \sum_{q_x,q_y,\nu=\pm} \left\{ \f{}{} \right.
  \Anu : |\left( \rhoonep + \nu \rhotwop \right)|^2: \nonumber \\
  &  &
  + \Bnu : |\left( \rhothrp + \nu \rhofoup \right)|^2: \\
  &  & + \VXY : \left[ \left( \rhoonep + \rhotwop \right) \right. \nonumber \\
  &  &    \left. \left. \times \left( \rhothrp + \rhofoup \right)^\dagger + h.c. \right]: \right\} \nonumber
\eea
where we have used $B$ to denote the function $A$ when its argument is $q_x$, and $\VXY=\VYX$.
For notational ease we have omitted the superscript $\alpha$ on the coupling constants.
This is the most general Hamiltonian bilinear in density operators consistent with the
symmetries of the square ($\Anu$ can also depend on $q_x$ generally, but this is omitted
here for simplicity).

It is convenient to introduce the combinations
\bea
   \rho_1 \pm \rho_2 & = & \sqrt{\f{L| q_x|}{2\pi}}
                             \left( \theta(q_x) \pm \theta(-q_x) \right)
                             \left( \ada \pm \alo \right) \nonumber \\
   \rho_3 \pm \rho_4 & = & \sqrt{ \f{L| q_y|}{2\pi} }
                             \left( \theta(q_y) \pm \theta(-q_y) \right)
                             \left( \bda \pm \blo \right) \nonumber \\
\eea
where $\theta$ is the step-function and $a$ and $b$ are commuting boson annihilation operators.
In order to diagonalize the Hamiltonian we express the $a$ and $b$ operators
in terms of new operators $\alpha$ and $\beta$ as follows
\bea
   \ada \pm \alo & = & e^{\mp \theta} \cos{\gamma}  \left( \alpd \pm \alp \right) \nonumber \\
                 &   &  +e^{\pm \phi} \sin{\gamma} \left( \betd \pm \bet \right) \\
   \bda \pm \blo & = & e^{\mp \theta^\prime} \cos{\gamma} \left( \betd \pm \bet \right) \nonumber \\
                 &   &  -e^{\pm \phi^\prime} \sin{\gamma} \left( \alpd \pm \alp \right)
\eea
where $\gamma,\theta,\theta^\prime,\phi,\phi^\prime$ are real numbers.
It is convenient to shorten the notation somewhat writing
${\cal A}^\pm = A^\pm_{q_y} L |q_x| / (2\pi)$, ${\cal B}^\pm = B^\pm_{q_x} L |q_y| / (2\pi)$ and ${\cal V} =V_{q_x,q_y} L \sqrt{|q_x| |q_y|}/(2\pi)$.

The $\alpha$'s and
$\beta$'s commute with each other. The preservation of the commutation relations of the $a$'s and $b$'s
imply the condition
\be
	\theta+\theta^\prime+\phi+\phi^\prime =0
\ee
By inserting the above expression into the Hamiltonian one finds a term proportional to
$(\alpha^\dagger-\alpha)(\beta^\dagger-\beta)$. The coefficient of this term can be
set equal to zero by setting
\be
	e^{2(\phi+\theta^\prime)} = \f{\cal{A}^-}{\cal{B}^-}.
\ee
Similarly there is a $(\alpha^\dagger+\alpha)(\beta^\dagger+\beta)$ term which coefficient
can be set to zero by choosing the angle $\gamma$ to fulfill
\be
     \tan (2 \gamma) =
     \f{{2 |\cal V}| \sqrt{ {\cal A}^- {\cal B}^- }}
       {
	 {\cal B}^+ | {\cal B}^- | - {\cal A}^+ |{\cal A}^- |
       }.
\ee
There are also $\alpha^\dagger \alpha^\dagger$ terms (similarly for $\beta$), getting rid of these fixes the angles
\bea
	e^{2\phi}  & = &\sqrt{ \f{ ({\cal A}^-)^2 (1-\tan^2\gamma)}{ {\cal B}^+ {\cal B}^- - {\cal A}^+ {\cal A}^- \tan^2 \gamma}}, \nonumber \\
	e^{2\theta^\prime} & = & \f{ {\cal A}^- }{ {\cal B}^- } e^{-2\phi}, \nonumber \\
	e^{2\phi^\prime}  & = &\sqrt{ \f{ ({\cal B}^-)^2 (1-\tan^2\gamma)}{ {\cal A}^+ {\cal A}^- - {\cal B}^+ {\cal B}^- \tan^2 \gamma}}, \nonumber \\
	e^{2\theta} & = & \f{ {\cal B}^- }{ {\cal A}^- } e^{-2\phi^\prime}. \nonumber
\eea
Having fixed the angles the Hamiltonian is diagonalized and reads
\be
H = \f{1}{2NL} \sum_{q_x,q_y} \left\{
	h_{\alpha} \alpha^\dagger \alpha
	+ h_{\beta} \beta^\dagger \beta + h.c. \nonumber
	\right\},
\ee
where
\be
   h_\alpha =
   	g_\alpha
	+\left( {\cal A}^+ {\cal A}^- \cos^2 \gamma + {\cal B}^+ {\cal B}^- \sin^2 \gamma \right)
	/g_\alpha,
\ee
and the parameter $g_\alpha = \sqrt{\f{ {\cal A}^+ {\cal A}^- - {\cal B}^+ {\cal B}^- \tan^2 \gamma}{1-\tan^2 \gamma}}$.
$h_\beta$ is gotten from the expression for $h_\alpha$ by interchanging ${\cal A}$s and ${\cal B}$s.

The asymptotic behavior of correlation functions are governed by small momentum values.
For $|q_x| \ll |q_y|$ we find
\bea
	\gamma & \rightarrow & V_{q_x=0,q_y} \f{\sqrt{A^-_{q_y} B^-_0}}{B^+_0 B^-_0} \f{|q_x|}{|q_y|},
		\nonumber \\
	e^{2\phi} & \rightarrow & \sqrt{ \f{(A_{q_y}^-)^2}{B^+_0 B^-_0}} \f{|q_x|}{|q_y|},
		\nonumber \\
	e^{2\theta^\prime} & \rightarrow & \sqrt{\f{B^+_0}{B^-_0}}, \nonumber \\
	e^{2\phi^\prime} & \rightarrow & \sqrt{\f{(B^-_0)^2}{A^-_{q_y} A^+_{q_y}}}
		\f{1}{\sqrt{1-\f{V_{0,q_y}^2}{A^+_{q_y} B_0^+}}} \f{|q_y|}{|q_x|}, \nonumber \\
	e^{2\theta} & \rightarrow & \sqrt{\f{A^+_{q_y}}{A^-_{q_y}}}
	\sqrt{ 1-\f{V_{0,q_y}^2}{A^+_{q_y} B_0^+}}, \nonumber \\
	h_\alpha & = & \f{L}{\pi} |q_x| \sqrt{ \f{A^+_{q_y} A^-_{q_y}}{1-\f{V_{0,q_y}^2}{A^+_{q_y} B_0^+}}},
	\nonumber \\
	h_\beta & = & \f{L}{\pi} |q_y| \sqrt{ B^+_{0} B^-_{0}}.
	\nonumber
\eea
The behavior for $|q_y| \ll |q_x|$ can be found by interchanging $A \leftrightarrow B$,$q_x \leftrightarrow q_y$ and primed $\leftrightarrow$ unprimed quantities.

\section{Operator dimensions}
We will now consider the stability of the square Fermi surface fixed point to products
of fermion chain operators on equal and opposite faces:
\be
	{\cal O} =\psi^{D_1}_{p_1 \sigma_1 l_1} \psi^{D_2}_{p_2 \sigma_2 l_2} \dots
	\psi^{D_M}_{p_M \sigma_M l_M}
\ee
all acting at the same $x$ and $\tau$, and where $D = +1$ means a creation
operator and $-$ means an annihilation operator, $p=+1,-1$ corresponds to
face 1 and 2 respectively and $\sigma = \pm 1$ denotes the spin direction.
To write the expression for the operator dimension of the above operator in a compact
form, we introduce four integer parameters,
$q_{\theta_c l} = \sum_{m=1}^M \delta_{l,l_m} D_m$,
$q_{\phi_c   l} = \sum_{m=1}^M \delta_{l,l_m} p_m D_m$,
$q_{\theta_s l} = \sum_{m=1}^M \delta_{l,l_m} \sigma_m D_m$,
and $q_{\phi_s   l}  =  \sum_{m=1}^M \delta_{l,l_m} p_m \sigma_m D_m$.
The dimension $d$ defined by $\langle {\cal O}(x) {\cal O}(0) \rangle = 1/x^{2d}$ is then
\bea \label{dimension}
	d & = &\f{1}{4} \f{1}{2\pi} \int_{-\pi}^{\pi} d \kper
	\f{1}{2} \sum_{l l^\prime} e^{i \kper \left( l - l^\prime \right)}
	\left( K^{-1}_{c \kper} q_{\theta_c l} q_{\theta_c l^\prime} \right. \\
&  & \left.              +K_{c \kper} q_{\phi_c l} q_{\phi_c l^\prime}
	      +K^{-1}_{s \kper} q_{\theta_s l} q_{\theta_s l^\prime}
              +K_{s \kper} q_{\phi_s l} q_{\phi_s l^\prime}
	\right) \nonumber
\eea
where $l,l^\prime$ are integers labeling the chains such that neighboring chains differ
by unity. The sum runs over all the chains present in the operator.
The Luttinger parameter which depends on the perpendicular momentum is given
by
\bea
	K_{\kper} & = & \left[ \cos^2(\gamma) e^{2\theta} + \sin^2(\gamma) e^{-2\phi} \right]_{\kpar \to 0}
		\nonumber \\
		&  = & \sqrt{\f{A^-_{\kper}}{A^+_{\kper}}} \f{1}{\sqrt{1-\f{V_{0,\kper}^2}{A^+_{\kper} B_0^+}}}.
\eea
This expression for the Luttinger parameter was also found in the context of crossed sliding Luttinger liquids\cite{Crossed}.

\begin{table}
\caption{\label{Table} Table of operator dimensions. The dimension is  $d=\int_{-\pi}^\pi \f{dk}{2\pi} d(k)$ where $d(k)$ is gotten from the table. The abbreviations $\cp = 1+\cos(k l)$ and $\cm = 1-\cos(k l)$. An asterisk $^*$ indicates that the operator does not correspond to a term in the microscopic Hubbard model.
}
\begin{ruledtabular}
\begin{tabular}{rll}
\# & Operator & $d(k)$ \\
1& $ \psi^\dagger_{1i\up} \psi_{1j\up}$ & $\f{1}{4}(K_c+K_c^{-1})\cm+ \f{1}{4} (K_s+K_s^{-1})\cm $\\
2&$\psi^\dagger_{1i\up} \psi^\dagger_{1i\down} \psi_{1j\down} \psi_{1j\up}$  &  $(K_c+K_c^{-1})\cm$  \\
3&$\psi^\dagger_{1i\up} \psi^\dagger_{1j\down} \psi_{1i\down} \psi_{1j\up} $ &   $(K_s+K_s^{-1})\cm$  \\
$4^*$&$\psi^\dagger_{1i\up} \psi^\dagger_{2i\up} \psi_{1j\up} \psi_{2j\up}$  &  $ K_c^{-1}\cm+ K_s^{-1}\cm$  \\
5&$\psi^\dagger_{1i\up} \psi^\dagger_{2i\down} \psi_{1j\down} \psi_{2j\up}$  &  $ K_c^{-1}\cm+ K_s \cp$  \\
6&$\psi^\dagger_{1i\up} \psi^\dagger_{2i\down} \psi_{1j\up} \psi_{2j\down}$  &  $ K_c^{-1}\cm+ K_s \cm$  \\
7&$\psi^\dagger_{1i\up} \psi^\dagger_{2j\down} \psi_{1i\down} \psi_{2j\up}$  &  $K_s^{-1} \cm + K_s \cp$  \\
$8^*$&$\psi^\dagger_{1i\up} \psi^\dagger_{2j\up} \psi_{1j\up} \psi_{2i\up}$  &  $K_c \cm +K_s \cm$  \\
9&$\psi^\dagger_{1i\up} \psi^\dagger_{2j\down} \psi_{1j\down} \psi_{2i\up}$  &  $K_c \cm +K_s \cp$  \\
10& $\psi^\dagger_{1i\up} \psi^\dagger_{2j\down} \psi_{1j\up} \psi_{2i\down}$  &  $K_c \cm +K_s^{-1}\cm$  \\
 11&$\psi^\dagger_{1i\up} \psi^\dagger_{1i\down} \psi_{2j\down} \psi_{2j\up}$  &   $K_c^{-1}\cm+K_c \cp $\\
 $12^*$&$\psi^\dagger_{1i\up} \psi^\dagger_{1j\up} \psi_{2i\up} \psi_{2j\up}$  &   $K_c \cp+K_s \cp$  \\
13&$\psi^\dagger_{1i\up} \psi^\dagger_{1j\down} \psi_{2i\down} \psi_{2j\up}$  &   $K_c \cp+K_s^{-1} \cm$  \\  \
14&$\psi^\dagger_{1i\up} \psi^\dagger_{1j\down} \psi_{2i\up} \psi_{2j\down}$  &   $K_c \cp+K_s \cm$
\end{tabular}
\end{ruledtabular}
\end{table}

\begin{table}
\caption{\label{Table2} Dimensions for operators occurring in various response functions. The dimension is  $d=\int_{-\pi}^\pi \f{dk}{2\pi} d(k)$ where $d(k)$ is gotten from the table. The abbreviations are as in Table~\ref{Table}. }
\begin{ruledtabular}
\begin{tabular}{rll}
type & Operator & $d(k)$ \\
${\cal O}_{\rm ss}$& $ \psi^\dagger_{1i\up} \psi^\dagger_{2j\down}$ & $\f{1}{4} \left(
K_c^{-1} \cp+ K_c \cm + K_s^{-1} \cm + K_s \cp \right)$\\
${\cal O}_{\rm ts}$& $ \psi^\dagger_{1i\up} \psi^\dagger_{2j\up}$ & $\f{1}{4} \left(
K_c^{-1} \cp+ K_c \cm + K_s^{-1} \cp + K_s \cm \right)$\\
CDW & $ \psi^\dagger_{1i\up} \psi_{2j\up}$ & $\f{1}{4} \left(
K_c^{-1} \cm+ K_c \cp + K_s^{-1} \cm + K_s \cp \right)$\\
SDW & $ \psi^\dagger_{1i\up} \psi_{2j\down}$ & $\f{1}{4} \left(
K_c^{-1} \cm+ K_c \cp + K_s^{-1} \cp + K_s \cm \right)$\\
    & $ \psi^\dagger_{1i\up} \psi^\dagger_{1j\up}$ & $\f{1}{4} \left(
K_c^{-1} \cp+ K_c \cp + K_s^{-1} \cp + K_s \cp \right)$\\
    & $ \psi^\dagger_{1i\up} \psi^\dagger_{1j\down}$ & $\f{1}{4} \left(
K_c^{-1} \cp+ K_c \cp + K_s^{-1} \cm + K_s \cm \right)$\\
    & $ \psi^\dagger_{1i\up} \psi_{1j\down}$ & $\f{1}{4} \left(
K_c^{-1} \cm+ K_c \cm + K_s^{-1} \cp + K_s \cp \right)$\\
\end{tabular}
\end{ruledtabular}
\end{table}

Specializing to the Hubbard Hamiltonian Eqs.~(\ref{Hubbardparams}), the Luttinger parameter in the charge sector $K_c$ can be written
\be \label{KcHubbard}
	K_c(\kper) = \f{1}{\sqrt{1+\f{\dlU}{1+\dlU f^2(0)}f^2(\kper)}}
\ee
where $\dlU = \f{Ua}{\pi v_0}$, and $f^2(0)=1/4$. the similar expression for $K_s$ differs by having $\dlU \to -\dlU$.
The change introduced by the adjacent face couplings is thus to renormalize the dimensionless coupling $\dlU \to \dlU/(1+\dlU f^2(0))$. While this is a small effect for small $\dlU$, it is very important for large $\dlU$ in that it isn't
possible to lower the value of $K_c$ arbitrarily much by increasing $\dlU$. In fact the lowest
possible value of $K_c$ is $1/\sqrt{2}$ which is reached for $\kper = 0$ at $\dlU \to \infty$.
For $\kper \to \pi$ the value of $K_c$ approaches 1 for all values of $\dlU$.

Table \ref{Table} shows the operator dimensions of all spin-conserving two and four-fermion interactions on same and opposite faces. Operators that reduce to pure density operators are not written down. A dimension smaller than 2 implies that the operator is relevant.
Similarly Table \ref{Table2} shows the operator dimensions of operators occurring in various response functions including singlet (${\cal O}_{\rm ss}$) and triplet ( ${\cal O}_{\rm ts}$) superconductivity as well as charge and spin density waves (CDW and SDW respectively).  

\section{Relevant operators}
While $K_s$ is different from unity for the Hubbard interaction, the SU(2) symmetry in spin
space requires $K_s$ to flow to $1$ as long as the spin-correlators decrease algebraically.
However if the system develops a spin gap the correlators decay no longer algebraically and
the spin sector is not part of the low energy physics.
Thus we consider these two cases separately.

\subsection*{a) No spin gap}

For the square Fermi surface to be stable upon doping, the operator
$\psi^\dagger_{i 1 \up} \psi_{j 1 \up}$ which comes from the chemical potential should in particular be irrelevant,
that is its dimension should be greater than 2. Using the expression Eq.~(\ref{dimension}) inserting
$K_s=1$ we find that the condition for irrelevancy is
\be \label{condition}
	\f{1}{4} \int_{-\pi}^{\pi} \f{dk}{2\pi}
 	\left( K_c(k)+K_c^{-1}(k) +2\right) (1-\cos(k(i-j))) > 2.
\ee
For momentum independent $K_c$ it
is easy to see that the chemical potential term is relevant for $ 3-2\sqrt{2} < K_c < 3+2\sqrt{2}$.
Taking into account the weak momentum dependency given by $f^2(k)$ and focusing on the nearest neighbor chains such that $i-j=1$ we find numerically that
the chemical potential term is irrelevant for $\dlU/(1+\dlU /4) > 1378.76$.
This inequality is clearly not fulfilled for any positive values of $U$, thus we conclude
that the square Fermi surface is unstable to Hubbard-like interactions upon doping.
This is at odds with the conclusions arrived at in Refs.\cite{Luther,Sudbo} where
it was found that a sufficiently large value of $U$ would make
the chemical potential term irrelevant. However there adjacent face couplings were not taken into account.

\begin{figure}
\includegraphics[clip,width=8cm]{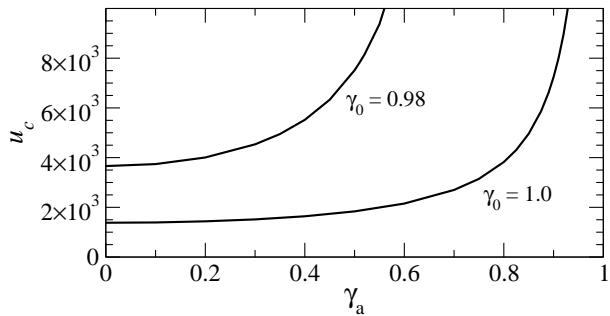}
\caption{$\dlU_c$ as a function of the anisotropy parameter $\gamma_a$ for two values of $\gamma_0$. For $\dlU > \dlU_c$, the chemical potential term is irrelevant and the square fermi surface is stable upon doping. \label{limits}}
\end{figure}

It is interesting to investigate what the anisotropies between the different face scattering rates must be
in order for the square to remain stable. Introducing anisotropy parameters describing the ratio of opposite to same face interactions $\gamma_o=U_{12}/U_{11}$, and adjacent to same face interactions
$\gamma_a=U_{13}/U_{11}$ we can write
\be
K_{c \kper}  =  \sqrt{ \f{\left[1+\dlU\f{1-\gamma_o}{2}f^2(\kper) \right]
          \left[1+\dlU \f{1+\gamma_o}{8} \right]}
	{ \left[1+\dlU \f{1+\gamma_o}{2} f^2(\kper) \right]
	  \left[1+\dlU \f{1+\gamma_o}{8} \right]
	-\f{\dlU^2}{4} \gamma_a^2 f^2(\kper)} }
\ee
where we have used $f^2(0)=1/4$.
For small $\dlU$, $K_{c \kper}$ is close to unity and the inequality (\ref{condition}) is violated.
Thus this stability
criterion sets a lower limit $\dlU_c$ on $\dlU$. For big $\dlU$, $K_c \propto \sqrt{1-\gamma_o}$, thus $\gamma_o$ should
be close to 1 in order to get a small $K_c$ which can saturate the inequality. How close can be found by setting $\dlU \to \infty$. It follows that $\gamma_o > 2\sqrt{2}/3 \approx 0.9428$. We have used the fact that $\gamma_a$ generally increases $K_c$, thus the lowest $U_c$ occurs for $\gamma_a=0$.
In Fig.\ref{limits} $\dlU_c$ is plotted as
a function of $\gamma_a$ for two values of $\gamma_0$. Note the extreme values of $\dlU$ needed for the square to remain
stable regardless of the anisotropy. It is unclear
how such extreme values can emerge in real systems of strongly correlated electrons.

The chemical potential term is not the only relevant operator when $K_c$ behaves according to Eq.~(\ref{KcHubbard}) taken from the microscopic Hubbard model.
As seen from Table \ref{Table}
other relevant operators are the back-scattering interactions (8)-(10) and umklapp processes (12)-(14).
There are also interactions that affect the spin sector only:  Operators
(3) and (7) can be interpreted to arise from interactions between spin variables on same/opposite faces and different chains. In the isotropic case $K_s=1$ they are both marginal operators.

\subsection*{b) Spin gap}
When the system develops a spin gap, spin correlators do not decay
as power-laws and the value of $K_s$ is not fixed to unity anymore.
Formally this situation can be covered by setting $K_s \to 0$ or $\infty$ dependent on
which field is pinned down. In either of these cases, because the chemical potential term involves both $K_s$ and $K_s^{-1}$ it
will automatically be irrelevant\cite{Zachar}. The physical reason for this is that removing a particle from a chain involves breaking the spin gap, and is therefore not part of the low-energy
physics. In the case where $K_s \to 0$ the operators (5),(6) corresponding to singlet pair-hopping, (8),(9) to different chain back-scattering, and
(12),(14) umklapp interactions are all potentially relevant dependent on the expression for $K_c$. On the other hand
when $K_s \to \infty$ the operators (4) corresponding to triplet pair-hopping, (10) to different chain back-scattering and (13) to umklapp scattering are potentially relevant.
The value of $K_s$ gotten from the microscopic Hubbard model indicates that if a spin gap forms it should be of the form $K_s \to \infty$. Using this together
with Eq.~(\ref{KcHubbard}) operators (4),(10) and (13) are all found to be relevant when considering nearest-neighbor chains.

The operators (2) corresponding to same-direction pair hopping and (11) umklapp scattering do not depend on $K_s$. With $K_c$ following Eq.~(\ref{KcHubbard}) operator (2) is found to be irrelevant while (11) is relevant for nearest-neighbor chains. When taking also inot account operators that describe interactions between chains at bigger separations ($|i-j|>1$) it becomes clear that there are a number of potential instabilities and a careful
renormalization group treatment which depends on the details of the microscopic Hamiltonian is needed to determine the dominant instability.

\section{Conclusion}
The basic insight gotten from taking adjacent face interactions into account in the boson Hamiltonian is  that it is not possible to make $K_c$ deviate significantly from unity even for
large values of the Hubbard-U interaction. Thus the ultimate fate of the system depends to a large extent on what happens in the spin sector.  If no spin gap develops, the square
Fermi surface is unstable upon doping unless there is a significant anisotropy between adjacent and same face scattering processes and the value of $U$ is very
large. On the other hand if a spin gap develops the chemical potential term is irrelevant.  However in that case there are also a number of other relevant operators which makes the dominant instability dependent on the details of the microscopic Hamiltonian.
 



\end{document}